\documentclass[prd,amssymb,aps]{revtex4}

\input graphicx
\input epsf
\input psfrag

\begin{document}

\newcommand{\myfrac}[2]{\mbox{$\textstyle\frac{#1}{#2}$}}

\title{Higher-derivative gravity in brane world models}

\author{M. Parry, S. Pichler and D. Deeg}

\affiliation{Arnold Sommerfeld Center, Department f\"ur Physik,
Ludwig-Maximilians-Universit\"at M\"unchen, Theresienstra{\ss}e 37,
D-80333 M\"unchen, Germany}

\begin{abstract}
We investigate brane world models in higher-derivative gravity theories where the gravitational Lagrangian is an arbitrary function of the Ricci 
scalar. Making
use of the conformal equivalence of such gravity models and
Einstein-Hilbert gravity with a scalar field, we deduce
the main features of higher-derivative gravity brane worlds. We
solve for a gravity model that has corrections quadratic in the
Ricci scalar and show one can evade both fine-tuning and the need 
for a bulk cosmological constant. An analysis of tensor and scalar 
perturbations shows gravity is localized 
on the brane and we recover the Newtonian limit.
\end{abstract}

\maketitle

\preprint{} LMU-ASC 08/05

\section{Introduction}

Brane world scenarios envisage our universe as a hypersurface embedded in
a higher-dimensional bulk. Large extra dimensions will not be visible in
collider experiments if standard model fields are confined to this
hypersurface. However, one should not expect gravity to be confined in
the same way. Randall-Sundrum (RS) models \cite{Randall:1999ee,key-3} are
simple self-consistent scenarios in which gravity {\em is} effectively
localized on a codimension one brane and the observed four-dimensional
Newtonian limit is reproduced. One also recovers standard Friedman
cosmology on the brane for energies low compared to the brane tension.
The key is the non-factorizable, warped metric of the bulk.

Nevertheless, RS models have an undesirable and much 
discussed
feature: the problem of fine-tuning. For the RS II
set-up which features one brane with positive tension $\lambda $ and a 
bulk cosmological constant $\Lambda $, we require
\begin{equation}\label{finetune}
\Lambda =-\myfrac{1}{6} \kappa _{5}^{2}\lambda^2.
\end{equation}
This fine-tuning is also necessary in cosmological models to obtain the
correct expansion behaviour of the observed universe 
\cite{Binetruy:1999hy}.
One of our goals here is to find ways to evade fine-tuning in brane world
models and without needing a bulk cosmological constant.

Since brane world models (mis)appropriate ideas from string/M-theory, it
is usually hoped that they will arise as effective models in some string
theory limit. From a phenomenological perspective therefore, one is led to
include other fields, for example moduli fields, or to consider
modifications to Einstein-Hilbert gravity. Bulk scalar fields have been
extensively considered because they can stabilize two-brane models and
also give rise to cosmological solutions. Because the Gauss-Bonnet term is
nontrivial in five dimensions, the consequences of adding this term to the
bulk action have also been explored.

Here we combine ideas from both these extensions of the usual
RS model. To be precise, we consider a gravitational
Lagrangian which is nonlinear in the Ricci scalar:
\begin{equation}
\label{higherderivativeaction}
S=-\frac{1}{2\kappa_n^2}\int d^{n}x\sqrt{|g|}f(R).
\end{equation}
We call this the physical or higher-derivative (HD) frame. In general,
such an action leads to equations of motion of fourth order and there is
little hope to solve them. However, this action is conformally equivalent
to Einstein-Hilbert gravity with a bulk scalar field. In this scalar field
(SF) frame, it is possible to obtain solutions to the equations of motion
and to transform the results back to the physical frame.

The outline of this article is as follows: first, we establish the
conformal equivalence of the HD and SF frames, and place brane worlds
within this framework. We present the complete action including the
appropriate Gibbons-Hawking boundary term for $f(R)$-gravity. In section
\ref{bsoln}, we construct brane world solutions in the SF frame for
potentials derived from superpotentials and for simple quadratic
potentials. Finally, we perturb about these background solutions in order
to show that gravity remains localized on the brane.  Our methodology
draws heavily on our recent work on bulk scalar field solutions
\cite{parry:2004iy}. We also include an appendix which details a
calculation in the HD frame.

\section{Conformal equivalence\label{conformalequivalence}}

Higher-derivative gravity theories that depend only on the Ricci scalar
are conformally related to usual scalar-tensor theories of 
gravity\cite{whitt}.
Consider the conformal transformation of the metric in $n$ spacetime
dimensions,
\begin{equation}
\label{transformation}
g_{AB}\rightarrow e^{2\omega }g_{AB}\equiv e^{2\kappa _{n}\phi /\sqrt{(n-1)(n-2)}}g_{AB},
\end{equation}
where $\kappa _{n}^{2}$ is the gravitational constant of the full
spacetime. The scalar field $\phi$ acts like a dilaton field. Under this
transformation, (\ref{higherderivativeaction}) becomes the
Einstein-Hilbert action plus the usual scalar field action:
\begin{equation}
S=\int d^{n}x\sqrt{|g|}\left[ -\frac{1}{2\kappa_n^2}R+\myfrac{1}{2}g^{AB}\phi _{A}\phi _{B}-V(\phi )\right] .
\end{equation} 
The potential $V(\phi)$ contains the information about the original
gravitational Lagrangian. In fact it is related to the Legendre
transformation of $f(R)$. If we introduce $\psi=f'(R)$ as the conjugate
variable to $R$, then
\begin{equation}
V(\phi )=-\frac{1}{2\kappa _{n}^{2}}\psi ^{\frac{n}{2-n}}f_{L}(\psi ),
\end{equation}
where 
\begin{equation}
f_L(\psi)\equiv\psi R-f(R) \quad \mbox{and} \quad \psi=e^{\kappa_n \phi \sqrt{\frac{n-2}{n-1}}}. 
\end{equation}  
From the theory of Legendre transformations, we know that $f_{L}(\psi )$
is well defined and concave as long as $f(R)$ is concave. Then it will be
possible to perform the inverse Legendre transformation to go back to the
HD frame.

Brane worlds, by their very nature, require us to add surface terms to the
action. We introduce one codimension one brane with tension $\lambda $ as
in the original RS II scenario and neglect matter fields on the brane.
Then the brane action is
\begin{equation}
S_{b}=-\lambda \int d^{n-1}\sigma \sqrt{|\gamma |}.
\end{equation}
Under the conformal transformation (\ref{transformation}), this becomes
\begin{equation}
S_b=-\lambda \int d^{n-1}\sigma \sqrt{|\gamma |}U(\phi), 
\label{couplingequation}
\end{equation}
where $U(\phi )\equiv e^{-\kappa _{n}\phi \sqrt{\frac{n-1}{n-2}}}$.

To obtain consistent conditions on the boundary, we have to include the equivalent of a Gibbons-Hawking term to the surface action in the HD frame (see also \cite{Barvinsky:1995dp}),
\begin{equation}
S_{GH}=\frac{1}{\kappa _{n}^{2}}\int d^{n-1}\sigma \sqrt{|\gamma
  |}f'(R)[K]_{\pm },
\end{equation}
where $K$ is the trace of the extrinsic curvature and $[K]_{\pm
}=K_{+}-K_{-}.$ (We define the unit normal to the surface $n^A$ normal to 
point into the $+$ region.) Transformation of the
boundary term to the SF frame yields the familiar Gibbons-Hawking term.

The equations of motion in the SF frame are 
\begin{eqnarray}
G_{AB} & = & \kappa _{n}^{2}\left[ \phi _{A}\phi _{B}-g_{AB}\left( 
\myfrac{1}{2}g^{CD}\phi _{C}\phi _{D}-V(\phi )\right) \right] 
,\label{eom1}\\ \nabla ^{2}\phi  & = & -V'(\phi ),\label{eom2} 
\end{eqnarray}
subject to the junction conditions
\begin{eqnarray}
\left[ K_{AB}\right] _{\pm } & = & \myfrac{1}{n-2}\kappa _{n}^{2}\lambda
U(\phi )q_{AB} , \label{jc1}\\
n^{A}\left[ \phi _{A}\right] _{\pm } & = & -\lambda U'(\phi ),\label{jc2}
\end{eqnarray}
where $q_{AB}$ is the projected metric on the surface\footnote{In the ``mainly negative'' signature, $q_{AB}\equiv g_{AB}+n_{A}n_{B}$ and we define $K_{AB}=q_A{}^C\nabla_C n_B$ and $K=q^{AB}K_{AB}=\nabla^A n_A$.}. 

\subsection{Relationship between $f(R)$ and $V(\phi)$}

While it is straightforward to compute $V(\phi)$ given an $f(R)$, the 
resulting expression is, in general, rather unwieldy. However, in the 
situations we consider, $V=0$ on the brane. If we then 
assume that for small $R$, $f(R)\sim R$, i.e. there is 
no bulk cosmological constant, then as an application of the mean value 
theorem, $R=0$ on the brane as well. It follows that in the vicinity of 
the brane
\begin{equation}\label{relation}
V(\phi)=\myfrac{1}{2}M^2(\phi-\overline{\phi})^2,
\end{equation}
where
\begin{equation}
\overline{\phi}=\sqrt{\myfrac{n-1}{n-2}}\ln[f'(0)] \quad \mbox{and} \quad 
M^2 = 
-\myfrac{n-2}{2(n-1)}\cdot\frac{[f'(0)]^{(n-4)/(n-2)}}{f''(0)}.\label{relnparams}
\end{equation}

\section{Brane world solutions}\label{bsoln}

From now on, and for obvious reasons, we specialize to the case of $n=5$ 
dimensions. We make all quantities dimensionless by introducing the
characteristic length scale $L=\kappa_5^{2/3}$. Then $\phi\rightarrow
L^{-3/2}\,\phi,\,\lambda\rightarrow L^{-4}\,\lambda$, etc. Actually, with
a different rescaling, it is possible to absorb $\lambda$ as well, but
this will make any fine-tuning of the parameters less apparent. In
particular, we do not want the potential to depend on the brane tension.

In the SF frame, where we perform all our calculations, the 
metric ansatz is
\begin{equation}
\label{metric}
ds^{2}=e^{2X(y)}\eta _{\mu\nu}dx^{\mu}dx^{\nu}-dy^{2},
\end{equation}
with the brane considered fixed at $y=0$. For simplicity, 
$\mathbb{Z}_{2}$-symmetry is assumed; we will write down solutions 
only for $y>0$. We consider 
perturbations about this background in the next section.

It follows that the metric in the HD frame is
\begin{equation}
\label{metricHD}
ds^{2}=e^{2A(Y)}\eta _{\mu\nu}dx^{\mu}dx^{\nu}-dY^{2},
\end{equation}
where
\begin{equation}
A(Y)=X(y)-\myfrac{1}{2\sqrt{3}}\phi(y) \quad \mbox{and} \quad Y=\int_0^y 
e^{-\phi(y')/2\sqrt{3}} dy',
\end{equation}
so that $Y=0$ gives the position of the brane. Without loss of generality, 
we let $A(0)=0$, which implies 
$X(0)=\myfrac{1}{2\sqrt{3}}\phi(0)\equiv\myfrac{1}{2\sqrt{3}}\overline{\phi}$.

The equations for the brane-bulk system are
\begin{eqnarray}
-3 X_{yy} - 6 X_y^2 &=& \myfrac{1}{2} \phi_y^2 + V,\\
6 X_y^2 &=& \myfrac{1}{2} \phi_y^2 - V,\label{constraint}\\
\phi_{yy}+ 4 X_y \phi_y &=& V',
\end{eqnarray}
with junction conditions
\begin{equation}
X_y(0) = -\myfrac{1}{6}\lambda U(\overline{\phi}),\quad \phi_y(0) = 
\myfrac{1}{2}\lambda 
U'(\overline{\phi}),
\end{equation}
where we evaluate quantities as $y\downarrow 0$. It follows from the form 
of $U$ and the constraint equation 
(\ref{constraint}) that, as promised, $V(\overline{\phi})=0$.

\subsection{Superpotential approach}

Superpotentials, familiar ingredients of supergravity theories, can be 
usefully employed for purely mathematical reasons \cite{DeWolfe:1999cp,Flanagan:2001dy,brax}. 
Given 
a superpotential 
$W(\phi)$, the associated potential in five dimensions is
\begin{equation}
V= \myfrac{1}{8}W'^{2}-\myfrac{1}{6}W^{2}.
\end{equation}
(Smuggled 
into the definition of
$W$ is an overall factor of $\mbox{sgn}(y)$, which we need to ensure 
$\mathbb
{Z}_{2}$-symmetry across the brane, but this does not affect the
calculation of $V$.) The advantage of such scalar field potentials is that the second order
equations become first order and the constraint equation is immediately
satisfied. Unfortunately, it is not possible to avoid fine-tuning in this 
class of
models. Additionally, the transformation back to the HD frame leads 
to an $f(R)$
that requires extra effort to make physical sensible. Nevertheless, these 
models are useful
because we can write down exact solutions and they illustrate
general features of brane world models in higher-derivative gravity. It 
is 
also possible to
find models in which the effective bulk cosmological constant,
$\Lambda\equiv\frac{1}{2}f(0)$, vanishes.

The system of equations becomes
\begin{eqnarray}
X_y=-\myfrac{1}{6}W,&& \quad \phi_y=\myfrac{1}{2}W',\\ 
W(\overline{\phi})=\lambda e^{-2\overline{\phi}/\sqrt{3}},&&\quad
W'(\overline{\phi})=-\myfrac{2}{\sqrt{3}}\lambda 
e^{-2\overline{\phi}/\sqrt{3}}.
\end{eqnarray}
Now it is clear that $W=W(\phi;\alpha)$---where $\alpha$ 
stands for the 
parameters of the superpotential---is a function of one variable that must 
satisfy two constraints. Thus we 
are forced to conclude that $\lambda=\lambda(\alpha)$, i.e. fine-tuning 
cannot be avoided.

\begin{figure}
\begin{center}
\psfrag{a}[]{\small $-1$}
\psfrag{b}[]{\small $-0.5$}
\psfrag{c}[]{\small $0.5$}
\psfrag{d}[]{\small $1$}
\psfrag{e}[r]{\small $-0.5$}
\psfrag{f}[r]{\small $-1.0$}
\psfrag{g}[l]{\small $f(R)$}
\psfrag{h}[]{\small $R$}
\psfrag{l}[l]{\small $\Lambda$}
\psfrag{1}[]{\small $R_2$}
\psfrag{2}[]{\small $R_1$}
\includegraphics[width=10cm,height=6cm]{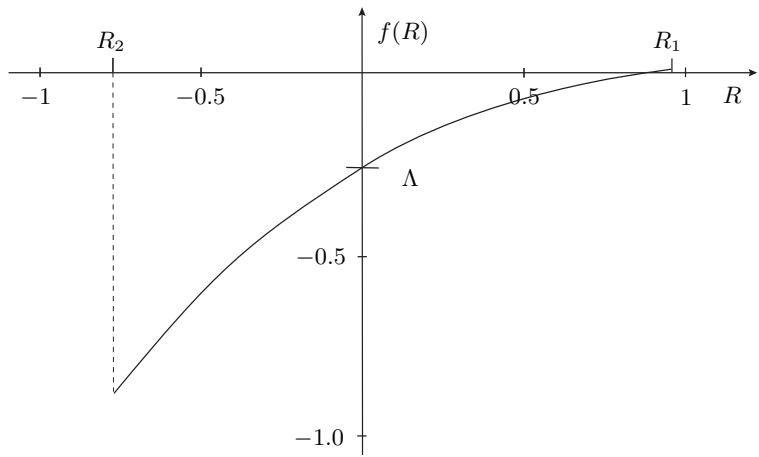}
{\caption{The gravitational Lagrangian $f(R)$ for a linear superpotential with 
$\alpha=-\myfrac{1}{2}\lambda$. The effective cosmological 
constant is negative for all $\alpha$.}\label{linearf}}
\end{center}
\end{figure}

\subsubsection{Linear superpotentials}

The superpotential $W(\phi )=2(\alpha \phi +\beta )$ leads to a quadratic
potential. After application of the junction conditions, we have
\begin{equation}
V(\phi) =-\myfrac{2}{3}\alpha^2 (\phi 
-\overline{\phi})\left(\phi-\overline{\phi}-\sqrt{3}\right),
\label{linearpotential} 
\end{equation}
where 
$\alpha=-\myfrac{1}{\sqrt{3}}\lambda e^{-2\overline{\phi}/\sqrt{3}}$. 
Note 
that the potential depends on $\lambda$ as expected and, because it is not 
of the form 
(\ref{relation}), we see immediately that the effective bulk cosmological 
constant cannot be 
zero.

From this potential, we can, in principle, compute $f_{L}(\psi )$, $R=
f'_L(\psi)$ and finally $f(R)$. However, the inverse Legendre
transformation does not exist when $f_L''(\psi)=0$. This occurs at two
points $\psi_1$ and $\psi_2$, corresponding to $R_1$ and $R_2$, where
$\psi_1<\overline{\psi}<\psi_2$ and $R_2<\overline{R}<0<R_1$. Because $R=f_L'$, as a function of $Y$, $R(Y)$ will remain in the interval $(R_2,R_1)$.

There are two options to make the theory physically sensible. First, we
could imagine $\min{\{R_1,|R_2|\}}>R_P=1$, so that the theory of gravity
as a whole would be modified before reaching these points. This is not so
attractive because already $|\overline{R}|\sim R_1,|R_2|$. The second
option is to assume that there is a second brane in the system at
$y=y_{\star}$, equivalently $Y=Y_{\star}$. Subject to appropriate boundary 
conditions,
\begin{equation}
X_y(y_{\star})=\myfrac{1}{6}\lambda_{\star}U(\phi_{\star}), \quad 
\phi_y(y_{\star})=-\myfrac{1}{2}\lambda_{\star}U'(\phi_{\star}),
\end{equation}
where quantities are evaluated as $y\uparrow y_{\star}$, this position
can be chosen so that $[\overline{\psi},\psi_{\star}]$ or
$[\psi_{\star},\overline{\psi}]$ are contained in $(\psi_1,\psi_2)$. 
The use of a so-called regulator brane to `slice off' singularities 
in a solution is well recognized in brane world models with scalar fields. 
Here, we are slicing off an offending piece of the theory space. In
this picture, the two branes and the higher-derivative theory of gravity
are inseparable; this is phenomenologically in the spirit of string
theory.

The expression for $f(R)$ in the invertible range is not very
illuminating; it is sketched in fig.(\ref{linearf}) for
$\alpha=-\myfrac{1}{2}\lambda$. However, it is not hard to show that 
$f(0)$ is
negative for all allowed values of $\alpha$. Therefore, as in the 
Randall-Sundrum case, the bulk cosmological constant is negative.

The solution in the SF frame is straightforward to 
obtain:
\begin{equation}
\phi(y)=\overline{\phi}+\alpha y,\quad
X(y)= 
\myfrac{1}{2\sqrt{3}}(\overline{\phi}+\alpha y)-\myfrac{1}{6}\alpha 
^{2}y^{2},
\end{equation}
so that
\begin{equation}
A(Y)=-2\left (\ln\left [1+\myfrac{1}{6}\lambda 
e^{-\sqrt{3}\,\overline{\phi}/2}Y\right]\right )^{2}.\label{linearsolution}
\end{equation}

Note that, in contrast to the original Randall-Sundrum case, $A(Y)$ is 
smooth across the brane. This is a corollary of the junction
conditions in the superpotential case and the form of $U$, but it is also
a generic feature of higher-derivative theories. Because the equations of
motion in the HD frame are fourth order, in general we expect 
discontinuities in
third order quantities. It is sometimes thought that the jump in the warp
factor is necessary to localize gravity on the brane, but we will show
this is not the case in the next section.

\subsubsection{Quadratic superpotentials}

The extra degree of freedom of a quadratic superpotential allows us to 
construct models in which the effective cosmological constant vanishes. 
Letting $W(\phi )=2(\alpha \phi ^{2}+\beta \phi +\gamma
)$ and adding the requirement that $f(0)=0$ to the junction conditions, 
we find
\begin{equation}
V(\phi)=-\myfrac{2}{3}\alpha^2 (\phi
-\overline{\phi})^2\left(\phi-\overline{\phi}-\sqrt{3}\right)^2,
\end{equation}
where now $\alpha=\myfrac{1}{3}\lambda e^{-2\overline{\phi}/\sqrt{3}}$. 
Near 
$\phi=\overline{\phi}$, we have $V=-2\alpha^2(\phi-\overline{\phi})^2$, 
which agrees with (\ref{relation}).

\begin{figure}
\begin{center}
\psfrag{a}[]{\small $-1$}
\psfrag{b}[]{\small $-0.5$}
\psfrag{c}[]{\small $0.5$}
\psfrag{d}[l]{\small $0.5$}
\psfrag{e}[l]{\small $-0.5$}
\psfrag{f}[l]{\small $f(R)$}
\psfrag{R}[]{\small $R$}
\psfrag{1}[]{\small $R_1$}
\psfrag{2}[]{\small $R_2$}
\includegraphics[width=10cm, height=6cm]{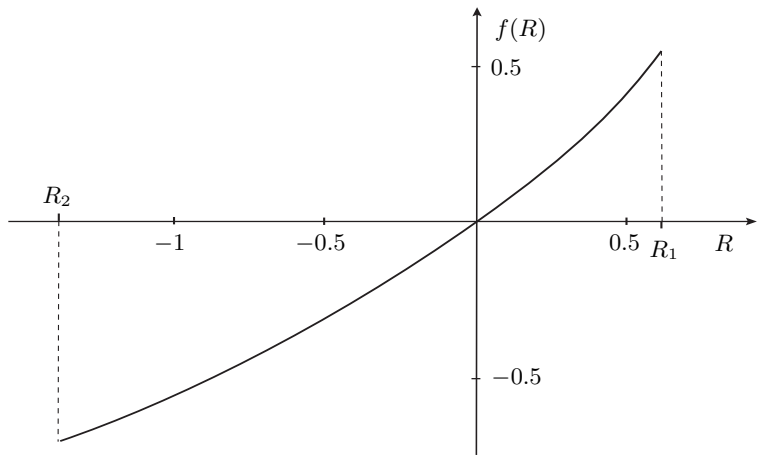}
{\caption{The gravitational Lagrangian $f(R)$ for a quadratic superpotential that leads to a vanishing effective cosmological constant. In this case 
$\alpha=\myfrac{1}{2}\lambda$.}\label{quadraticf}}
\end{center}
\end{figure}

As before, we have to be careful with the inverse Legendre transformation. The
situation is as with the linear superpotential case except that
$\overline{R}=0$. (Actually, $f_L''$ has additional zeros but there is only one interval $(\psi_1,\psi_2)$ within in which we obtain a self-consistent solution.) Once more the final expressions are a little unwieldy and we resort to graphical representation: fig.(\ref{quadraticf}) shows $f(R)$ when $\alpha=\myfrac{1}{2}\lambda$. There are $+R^2$ corrections to $f(R)$ near $R=0$.

\begin{figure}
\begin{center}
\psfrag{a}[]{\small $0$}
\psfrag{b}[]{\small $5$}
\psfrag{c}[]{\small $10$}
\psfrag{d}[r]{\small $0.5$}
\psfrag{e}[r]{\small $1$}
\psfrag{f}[r]{\small $e^{A(Y)}$}
\psfrag{Y}[]{\small $Y$}
\psfrag{h}[]{\small $-5$}
\psfrag{k}[]{\small $-10$}
\includegraphics[width=10cm, height=6cm]{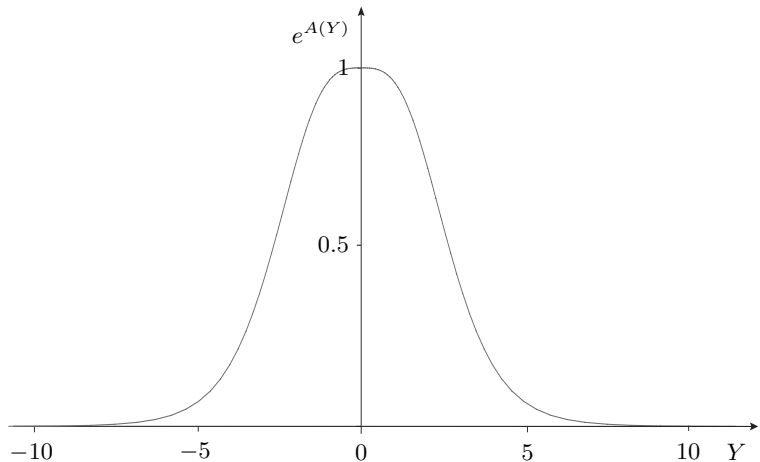}
{\caption{The warp factor $e^{A(Y)}$ for a quadratic 
superpotential where $\Lambda=0$ and $\alpha=\myfrac{1}{2}\lambda$.}\label{quadraticwarp}}
\end{center}
\end{figure}

The solutions for $\phi(y)$ and $X(y)$ in the SF frame can again be
straightforwardly written down and we omit the details. This time,
however, there is no closed form expression for $A(Y)$ in the HD frame.
We depict it graphically in fig.(\ref{quadraticwarp}), where it can be
seen to be very smooth across the brane.

\subsection{Simple quadratic potentials}

The superpotential models are easy to solve and illustrative. In
particular, we have learned that if $f(R)$ has a term quadratic in $R$
with positive coefficient, it appears we can avoid having a bulk
cosmological constant. To show this is indeed a general result, we 
consider the 
simple quadratic potential,
\begin{equation}
V(\phi )=\myfrac{1}{2}M^{2}\phi ^{2}.
\end{equation}

Using nonlinear perturbation theory, we recently \cite{parry:2004iy} found
brane world solutions for massive scalar fields in the limit of small
$M^{2}$ and large $-M^{2}$. (In terms of our scaling here, `small' and 
`large' mean with respect to $\lambda^2$.) Such potentials do not 
require 
fine-tuning,
but the price we pay is in singularities that now crop up a finite
distance from the brane. As indicated earlier, this generic feature of
bulk scalar field models is remedied by the inclusion of a regulator
brane.

Returning to the HD frame is once again problematic since $f_L$ is not 
concave everywhere---though where it breaks down is independent of $M$. We deduce 
from eq.(\ref{relnparams}), however, that 
in the region of invertibility
\begin{equation}\label{ftaylor}
f(R)=R-\myfrac{3}{16}M^{-2} R^{2}+\ldots.
\end{equation}
We indicate in the appendix that 
away from the brane this theory is quite different to a higher-derivative 
theory in which $f(R)$ has only $R^2$ corrections.

\subsubsection{Small $M^2$}

Following \cite{parry:2004iy}, we let $\epsilon=M^2$ be our perturbative 
parameter and introduce the strained coordinate $x=B^{-1}y$.
The zeroth order solutions in $x$ are 
$X_0=\myfrac{1}{2\sqrt{3}}\phi_0=\myfrac{1}{4}\ln(1-\myfrac{2}{3}\lambda 
x)$, indicating a singularity at $x=\myfrac{3}{2}\lambda^{-1}$. Omitting 
the details of the first order solutions $X_1$ and $\phi_1$, we find
\begin{equation}
A(Y)=-\myfrac{1}{4\sqrt{3}}M^2 \int_0^x dx 
\frac{\phi_0^2}{\phi_{0,x}}=-\myfrac{9}{128}\cdot 
\frac{M^2}{\lambda^2}\left[ 2u^2\ln u (\ln 
u-1) +u^2 -1\right],
\end{equation}
where $u=1-\myfrac{2}{3}\lambda x$. Additionally, we must have 
$B=1-\myfrac{1}{6}M^2/\lambda^2$ in order to avoid secular terms at first 
order. The coordinate $Y$ in the HD frame is approximately
\begin{equation}
Y=2\lambda^{-1}\cdot\myfrac{3B}{4-B}(1-u^{3/4}).
\end{equation}
Therefore, near the brane, $A(Y)=\myfrac{1}{36}\lambda M^2 
Y^3+\ldots$. This result is in perfect agreement with the HD frame  
calculation given in the appendix and shows $A(Y)$ will have 
discontinuities at 
the brane in its 
third derivatives. The sign of $M^2$ determines whether the warp factor 
increases or decreases away from the brane. Since we want the latter 
case, we take $M^2$ negative from now on.

The situation as we approach the singularity is quite remarkable. While 
$X$ and $\phi$ both diverge, $A$ is well behaved. To first order, and 
this is also borne out by numerical simulations, as 
$u\rightarrow 0$, $Y\rightarrow Y_s\simeq 2\lambda^{-1}$ and we have
\begin{equation}
A(Y_s)=\myfrac{9}{128}\cdot\frac{M^2}{\lambda^2}<0 \quad \mbox{and} \quad 
A'(Y_s)=0.
\end{equation}
Thus, while our solution in the SF frame fails at the singularity, it 
appears the solution in the HD frame will continue to $Y>Y_s$. Although 
we have been unable to perform the HD frame calculation to 
prove this, it is moot since we require a regulator brane at 
$Y_{\star}<Y_s$ to make the higher-derivative theory sensible.

\begin{figure}
\begin{center}
\psfrag{a}[r]{\small $0$}
\psfrag{b}[]{\small $1$}
\psfrag{c}[]{\small $2$}
\psfrag{d}[]{\small $3$}
\psfrag{e}[r]{\small $0.5$}
\psfrag{f}[r]{\small $1.0$}
\psfrag{g}[l]{\small $e^{A(Y)}$}
\psfrag{y}[]{\small $Y$}
\includegraphics[width=10cm,height=6cm,angle=0]{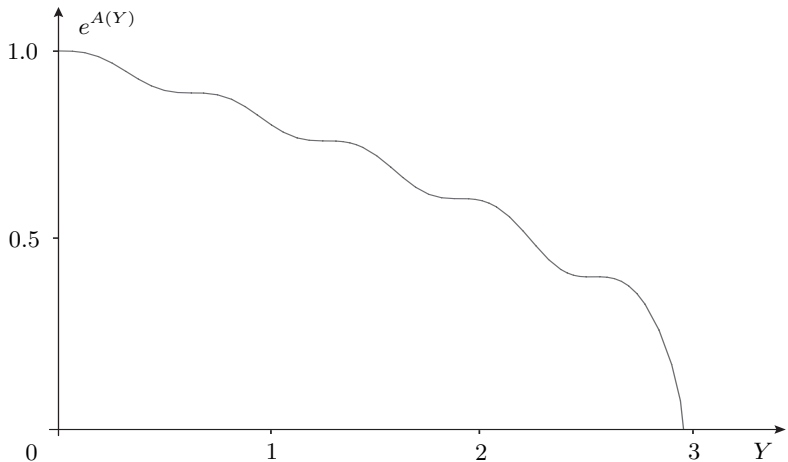}
{\caption{The warp factor $e^{A(y)}$ for a 
quadratic potential with $M^2=-100\lambda$.}\label{squarefigure}}
\end{center}
\end{figure}

\subsubsection{Large $-M^2$}

For large negative $M^2$, the appropriate perturbative parameter is 
$\epsilon=|M|^{-1}$. It turns out $X$ and $\phi$ functionally depend on 
both $y$ and $|M|y$. This means derivatives of these quantities are of 
{\em lower} order in perturbation theory. To first order, we find
\begin{equation}
X(y) = \myfrac{1}{2}\ln\left (1-\myfrac{1}{3}\lambda y\right ), \quad
\phi(y)= -\myfrac{1}{\sqrt{3}}\lambda |M|^{-1}\left 
(1-\myfrac{1}{3}\lambda y\right 
)^{-1}\sin(|M|y).
\end{equation}
There is a singularity as $y\rightarrow 3\lambda^{-1}$ but $\phi$ is 
first order 
so remains small and the higher-derivative theory sensible until 
quite close to the singularity. At second order, $X$ also picks up an 
oscillatory term, but $A$ has an oscillatory 
component already at first order:
\begin{equation}\label{HDlargeM}
A(Y)=\myfrac{1}{2}\ln\left( 1-\myfrac{1}{3}\lambda Y\right) 
+\myfrac{1}{6}\lambda |M|^{-1}\left(1-\myfrac{1}{3}\lambda 
Y\right)^{-1}\sin(|M|Y).
\end{equation}
This result is entirely consistent with the calculation in the HD frame 
outlined in the appendix. The smoothness of the warp factor across the 
brane 
and its oscillatory behaviour in the bulk is illustrated in 
fig.(\ref{squarefigure}) for the case $M^2=-100\lambda$.

As pointed out in \cite{parry:2004iy}, the large negative $M^2$ limit in 
the SF frame looks like the usual RS model with vanishing bulk cosmological constant {\em if} eq.(\ref{constraint}) were ignored. It is this equation which gives rise to the fine-tuning condition (\ref{finetune}). But the situation in the HD frame is even more curious: eq.(\ref{ftaylor}) suggests $f(R)\rightarrow R$. In other words, in what looks to be Einstein-Hilbert gravity, we seem to be able to embed a brane with non-zero tension into a bulk without a cosmological constant! Of course, appearances are deceptive. For a start, this solution has undesirable limiting behaviour: while $A\rightarrow A_0\equiv\myfrac{1}{2}\ln(1-\myfrac{1}{3}\lambda Y)$, $A_Y\nrightarrow A_{0,Y}$ etc. Second, $f(R)\rightarrow R$ only in a certain range of $R$; we need a further brane to make the theory sensible.

\section{Perturbations and the Newtonian limit on the brane}

The background solutions we have found for the higher-derivative theories
do not have a kink at the brane in the HD frame. While it is known that a kink is not a prerequisite for gravity to be localized on the
brane \cite{kofman2}, i.e. in order to have a zero mode, it is important to check that the brane-bound observer sees
Newtonian gravity. Building on \cite{parry:2004iy}, we
show that this is indeed the case in the framework of the simple
quadratic potentials presented in the last section.

For computational ease, we now choose a different gauge for the SF frame 
background: 
\begin{eqnarray}
ds^2=a^2(y)(\eta_{\mu\nu}dx^\mu dx^\nu - dy^2).
\end{eqnarray}
Scalar and tensor perturbations 
about this 
background are described by \cite{mfb}
\begin{eqnarray}
\delta g_{AB}=a^2(y) 
 \left( 
\begin{array}{ccc|c}
 & & &\\[-1.5ex]
 & 2\psi \eta_{\mu\nu}+ 2E_{,\mu\nu}+h_{\mu\nu} & & B_{,\mu}\\[-0.8ex]
 & & & \\
\cline{1-4} & & & \\[-2.2ex]
 & B_{,\mu} & & 2A \\
\end{array} \right) ,
\end{eqnarray}
where $h_{\mu\nu}$ is transverse and traceless, and indices are raised
(lowered) by $\eta^{\mu\nu}$ ($\eta_{\mu\nu}$). We have neglected vector
perturbations because they will not have support to linear order.

The tensor perturbations are already gauge invariant, and the pertinent 
gauge invariant scalar combinations are
\begin{eqnarray}
\mathcal{A}&=&A-(B-E_y)_y-\frac{a_y}{a}(B-E_y), \\
\Psi&=&\psi+\frac{a_y}{a}(B-E_y).
\end{eqnarray}
Transforming back to the HD frame \cite{mfb}, we 
find $a_{HD}=a\,e^{-\phi/2\sqrt{3}}$, $A_{HD}=A+\myfrac{1}{2\sqrt{3}}\delta\phi$, 
and $\psi_{HD}=\psi-\myfrac{1}{2\sqrt{3}}\delta\phi$. However, $B$, $E$ and 
$h_{\mu\nu}$ are invariant. The gauge invariant quantities in the HD 
frame are
\begin{eqnarray}
\mathcal{A}_{HD}&=&\mathcal{A}+\myfrac{1}{2\sqrt{3}}\delta\phi^{(gi)}, \\
\Psi_{HD}&=&\Psi-\myfrac{1}{2\sqrt{3}}\delta\phi^{(gi)},
\end{eqnarray}
where $\delta\phi^{(gi)}=\delta\phi+\phi_y(B-E_y)$ is the gauge 
invariant perturbation of $\phi$. Equivalently, we may write
\begin{equation}
\delta\phi^{(gi)}=\myfrac{2}{\sqrt{3}}\cdot[\ln{f'(R)}]'\,\delta 
R^{(gi)}.
\end{equation}
We adopt the 
brane world equivalent of 
the longitudinal gauge in which 
$B=E=0$ from now on.

In addition to the metric and scalar field perturbations, we must add a 
matter perturbation to the brane in the HD frame, $\mathcal{L}_{\delta 
m}=\mathcal{L}_{\delta m}(\gamma^{\mu\nu}_{HD})$. The relation 
between 
the stress-energy tensors in the two frames is then
\begin{equation}
\tau_{\mu\nu}=T(\phi)\tau^{HD}_{\mu\nu},
\end{equation}
where $T(\phi)=e^{-\phi/\sqrt{3}}$. Note that indices of the 
stress-energy tensor in the SF frame are raised 
(lowered) by $\gamma^{\mu\nu}$ ($\gamma_{\mu\nu}$), where 
$\gamma_{\mu\nu}=a^2\eta_{\mu\nu}$. It follows that $\tau=U(\phi) 
\tau_{HD}$. Actually, the distinction will not be important for us
because the $\phi$ appearing is the background $\phi$ on the brane and 
$\phi(0)=\overline{\phi}=0$ in our scenario. Therefore $a(0)=a_{HD}(0)=1$.

The final perturbation that must be considered is the perturbation in the
brane position. We suppose the brane is perturbed from $y=0$ to
$y=\zeta(x^{\mu})$. This scalar degree of freedom is manifestly gauge
invariant in the longitudinal gauge and leads to an additional term in
the perturbed junction conditions. We assume, however, that there is no
matter on the regulator brane at $y=y_{\star}$, nor is its position
perturbed.

\subsection{Tensor perturbations}

The tensor modes are the most important contribution to the Newtonian
limit on the brane. It was shown in \cite{parry:2004iy} that the Green's
function for $h_{\mu\nu}$ has the required $1/r$ behaviour and so we
will not labour the point here.

We found that there is always a zero mode of the perturbations
$h_{\mu\nu}$ which is proportional to $a^{3/2}$, and that there are no
tachyonic modes. Furthermore, for small $M^2$ and for large negative
$M^2$, massive modes of $h_{\mu\nu}$ only make small corrections to the
Green's function. The brane bending is given by
$\Box\zeta=-\myfrac{1}{6}\tau$, as first pointed out in
\cite{Garriga:1999yh}.

\subsection{Scalar perturbations}

We have to give special attention to scalar perturbations because
they affect the transformation back to the physical frame. The danger is that the Newtonian potential may pick up an additional
scalar mode in the HD frame. Our primary concern is the existence of zero and tachyonic modes. These  have been carefully studied by Kofman and Mukohyama \cite{Mukohyama:2001ks}, and Lesgourgues and Sorbo \cite{Lesgourgues:2003mi}.

The off-diagonal $[\mu,\nu]$-components of the scalar equations of 
motion imply
\begin{equation}
\mathcal{A}=2\Psi.
\end{equation}
The $[5,0]$-component is a constraint equation, 
\begin{eqnarray}
3\Psi_y + 6H\Psi  = -\phi_y\, \delta\phi^{(gi)},\label{05constraint}
\end{eqnarray}
where $H=a_y/a$, which we use to find $\delta\phi^{(gi)}$. The one 
independent scalar 
equation of motion can be written as a wave equation:
\begin{eqnarray}
\Psi_{yy}-\Box\Psi + \left( 3H-2\frac{\phi_{yy}}{\phi_y} \right) \Psi_y 
+4 \left( H_y-H\frac{\phi_{yy}}{\phi_y} \right) \Psi 
=0.\label{waveequation}
\end{eqnarray}

The junction conditions at $y=0$ are
\begin{eqnarray}
\delta\phi_y^{(gi)}(0) = \myfrac{1}{2}\lambda a \left(U'' 
\delta\phi^{(gi)}- 2 U' \Psi\right)+\myfrac{1}{4\sqrt{3}}\,a U\tau, \quad
\Psi_y(0) = -\myfrac{1}{6} \lambda a\left(U' \delta\phi^{(gi)}-2U 
\Psi\right).
\end{eqnarray}
The junction conditions at $y=y_{\star}$ are the same except for an overall sign 
and they do not include the term in $\tau$. This term comes from 
computing $\delta\mathcal{L}_{\delta m}/\delta\phi$ in the SF frame, and 
it prevents the scalar perturbations from vanishing completely. It also 
means we cannot fully adopt the machinery of \cite{Lesgourgues:2003mi}.

We solve equation (\ref{waveequation}) by introducing a 
nonsingular Mukhanov
variable. In our scenarios, $H$ is never zero, so the variable that
remains regular is $v= a^{3/2}(\delta\phi^{(gi)}-\phi_y\Psi/H)$. The
equation for $v$ is
\begin{equation}\label{ueom}
v_{yy}-\Box v-\frac{z_{yy}}{z}v=0,
\end{equation}
where $z=a^{3/2}\phi_y/H$. Since $v_y$ will have a jump at the brane 
positions, the equation for $v$ can profitably be recast as a Green's 
function problem. Following the same procedure in \cite{parry:2004iy}, 
the problem reduces to finding the normalized modes $\psi_m$ that satisfy
\begin{equation}\label{evalue}
(\hat{D}_+\hat{D}_- +m^2)\psi_m=0,
\end{equation}
where $\hat{D}_{\pm}=\partial_y\pm z_y/z$, subject to the boundary 
conditions $\partial_y\psi_m(0)=\partial_y\psi_m(y_{\star})=0$. It 
follows that
\begin{equation}
m^2\geq\left. 2\frac{z_y}{z}\psi_m^2\right|_0^{y_{\star}}.
\end{equation}
Unfortunately, our ignorance of $\psi_m$ at the branes prevents us from 
being able to compute the sign of the right hand side. We do know 
that $z_y/z$ becomes infinite if $\phi_y=0$, thus the existence of 
tachyonic modes cannot be ruled out. However, the arguments of 
\cite{Lesgourgues:2003mi} suggest we can avoid them in the cases we 
consider. To be precise, {\em if} we neglect the matter perturbation on 
the brane, then $\phi_y$ not zero on $[0,y_{\star}]$ implies\footnote{Technically, there are three additional criteria but they are met by our background solutions.} we 
have only modes with $m^2\geq 0$. We now show explicitly that there are 
no zero modes.

The general solution to eq.(\ref{evalue}) for $m^2=0$ is
\begin{equation}
\psi_0(y)=c_1 z(y)+c_2 z(y)\int_{y_{\star}}^{y}z^{-2}(y')dy',
\end{equation}
with $c_1,\,c_2$ constants. If the boundary conditions are satisfied only 
if $c_1=c_2=0$, there is no zero mode.

\subsubsection{Small $M^2$}

In this case $\phi_y$ has no zeros. To a very good approximation, the 
background solutions give
\begin{equation}
z(y)\simeq 2\sqrt{3}\left(1-\frac{\lambda y}{r}\right)^{r/4},
\end{equation}
where $r\simeq 2-\myfrac{4}{9}M^2/\lambda^2$ to first order. It is 
straightforward to check that for all values of $r$, $\psi_0$ must be 
zero.

\subsubsection{Large $-M^2$}

There is a zero almost immediately in $\phi_y$, namely at $y\simeq
\pi/2|M|$. To avoid this, the regulator brane must be brought very close
to the visible brane, i.e. $y_{\star}\sim |M|^{-1}$. This is exactly the 
requirement we found in
\cite{parry:2004iy}. Then
\begin{equation}
z(y)\simeq 2\sqrt{3}\left(1-\myfrac{1}{2}M^2 y^2\right),
\end{equation}
and once more $\psi_0=0$.

\section{Conclusions}

We have initiated a study of brane world models in a class of higher-derivative gravity theories that are conformally equivalent to Einstein-Hilbert gravity with a scalar field. For illustrative purposes, we started with scalar
field potentials that are derived from superpotentials. These exhibit the main
features of higher-derivative gravity in brane world models. Then, we considered
simple quadratic potentials which mimic the addition of $R^2$ terms to the gravitational Lagrangian. The typical background solutions are very smooth across the brane in the physical frame---in contrast to the usual scenarios. Furthermore, we found that we do not need a bulk cosmological constant and, therefore, can avoid the fine-tuning problem that plagues the 
original Randall-Sundrum scenario.  Finally, we were able to ensure that gravity
is effectively localized on the brane and that the Newtonian limit holds. Although the
equations of motion are more difficult to solve in higher-derivative
gravity theories, such theories offer an intriguing alternative to standard formulations
of brane world models. We are now working on the cosmological consequences of this approach.

\acknowledgments
It is a pleasure to thank Dani\`{e}le Steer and Alexander Vikman for stimulating discussions. MP and DD were supported by SFB 375.

\appendix

\section*{Appendix}

It is difficult to derive the equations of motion for $f(R)$ theories of
gravity, let alone to solve them. Here we present the simplest
higher-derivative theory,
\begin{equation}\label{HDaction}
f(R)=R+\sigma^2 R^2,
\end{equation}
specialized to the case of  $\mathbb{Z}_{2}$-symmetric brane worlds in five dimensions.

The line-element in the HD frame that will give us both the equation of motion and the constraint equation when we vary with respect to the metric functions is
\begin{equation}
ds^2=a^2(Y)\eta_{\mu\nu}dx^{\mu}dx^{\nu}-b^2(Y)dY^2.
\end{equation}
Afterwards, we can put $a(Y)=e^{A(y)}$ and $b(Y)=1$. In terms of dimensionless quantities, the total action is
\begin{equation}
S=-\myfrac{1}{2}\int d^4 x\, dY a^4 b \left(R+\sigma^2 R^2\right) + \int d^4 x\, a^4 \left[(1+2\sigma^2 R)[K]_{\pm} -\lambda\right],
\end{equation}
where
\begin{equation}
R=\frac{4}{b^2}\left(\frac{2a_{YY}}{a}+ \frac{3a_Y^2}{a^2}-\frac{2a_Y b_Y}{a b}\right) \quad \mbox{and} \quad [K]_{\pm} = -\left.\frac{8}{b}\cdot \frac{a_Y}{a}\right.
\end{equation}

The bulk equations are
\begin{eqnarray}
6A_Y^2 + 8\sigma^2\left(8A_{YYY} A_Y-4A_{YY}^2 +32A_{YY}A_Y^2+5A_Y^4 \right) &=& 0,\\
6A_Y^2 + 3 A_{YY} + 8\sigma^2\left(2A_{YYYY}+16A_{YYY}A_Y+12A_{YY}^2+37A_{YY}A_Y^2+5A_Y^4 \right) &=& 0.
\end{eqnarray} 
The junction conditions follow from taking the variations $\delta a,\,\delta a_Y,\,\delta a_{YY}$ etc to be independent on the brane. We find
\begin{eqnarray}
\sigma^2 A_Y(0) &=&0,\\
6 A_Y(0) + 32\sigma^2 A_{YYY}(0) &=&-\lambda,
\end{eqnarray}
and $A_{YY}(0)=0$ by $\mathbb{Z}_{2}$-symmetry. When $\sigma^2\neq 0$, 
$A_Y(0)=0$ and so there is a jump in the third derivative only. 
Substituting (\ref{HDaction}) into (\ref{relnparams}), we find $M^2=-\myfrac{3}{16}\sigma^{-2}$ and conclude $A(Y)=\myfrac{1}{36}\lambda M^2 Y^3$ in the vicinity of the brane. This confirms the result obtained in the SF frame. 

It is not possible to solve the fourth order equations in general. In the large $\sigma^2$ limit, numerical simulations show that $A$ decreases monotonically to a singularity at $Y=Y_s$, where $Y_s$ is an increasing function of $\sigma$. This is in stark contrast to the small $M^2$ limit away from the brane found in the text. On the other hand, for small $\sigma^2$, it is straightforward to show that (\ref{HDlargeM}) is indeed the approximate solution in the HD frame. We conclude that bulk singularities are likely to plague the higher-derivative theories considered here.

\end{document}